\input{aipcheck.tex}

\documentclass[]{aipproc}

\layoutstyle{6s}

\begin{document}

\title{DVCS @ HERMES}

\author{Morgan J. Murray for the H{\sc ermes} Collaboration}{
	address={SUPA, School of Physics \& Astronomy, University of Glasgow, Glasgow, G12 8QQ, Scotland}
}


\copyrightyear {2010}

\begin{abstract}
The study of Generalised Parton Distributions (GPDs) promises to provide new knowledge of the structure of the nucleon, including, most notably, access to the total angular momentum of quarks within the nucleon. It can be difficult to ascertain new information on the distributions, but amongst all the exclusive processes that can provide access, Deeply Virtual Compton Scattering (DVCS) is relatively simple and experimentally accessible. The H{\sc ermes} collaboration has the most diverse results pertaining to DVCS of any experiment, extracting asymmetries in the azimuthal distribution of produced photons according to both beam helicity and charge and target spin state. In this talk, we provide an overview of the H{\sc ermes} DVCS result catalogue and explain how the results are used to improve constraints on the underlying GPDs. \end{abstract}

\date{\today}
\maketitle
\section{Introduction}

The H{\sc ermes} experiment at {\sc desy} ran from 1995 to 2007 as a forward spectrometer on the H{\sc era} electron/positron beam. Its original \emph{raison d'\^etre} was the investigation of the spin structure of the nucleon, a purpose continued in the later years of the experiment's lifetime with its focus on exclusive physics. These proceedings report briefly on selected H{\sc ermes} results on Deeply Virtual Compton Scattering which is the leptoproduction of real photons from a nuclear target --- in the case of H{\sc ermes} this reduces to electroproduction on typically a Hydrogen target.

The study of exclusive physics is performed with a view to obtaining information on Generalised Parton Distributions (GPDs). These theoretical objects contain a wealth of information on nucleonic structure, including a way to access the total angular momentum of the constituent partons~\cite{Ji1997} and correlated information pertaining to the distribution of partons in the transverse spatial plane of the nucleon with the fraction of the nucleon's longitudinal momentum carried by that parton~\cite{BurkardtPhys.Rev.D62:0715032000;Erratum-ibid.D66:1199032002}. There are four distributions of interest that are expected to enter into typical scattering experiments at leading twist with the least kinematical suppression: $H, E, \widetilde{H}$ and $\widetilde{E}$. Generalised Parton Distributions only describe the soft part of the diagram and thus appear in data from lepton scattering experiments convoluted with a hard scattering kernel~\cite{Balitsky1997}. The resultant distribution is referred to a Compton Form Factor (CFF) and is typically denoted $\mathcal{H}, \mathcal{E},\mathcal{\widetilde{H}}$ and $\mathcal{\widetilde{E}}$ for GPDs $H, E, \widetilde{H}$ and $\widetilde{E}$ respectively.

The process $ep\rightarrow ep\gamma$ has two contributors; alongside the DVCS process there is a contribution from the Bethe Heitler process ($ep$ scattering with a Brehmsstrahlung photon). The scattering amplitudes from each process interfere and provide the cross-section with three contributory terms: one term from the squared amplitude from each process and one from the interference between the two processes. Typically at the kinematic space covered by H{\sc ermes} the BH process dominates but access to the large interference term can provide information derived form CFFs and thus GPDs. Mostly asymmetries are constructed so as to minimise the contribution from the pure BH process.

\section{The HERMES Experiment}
The H{\sc ermes} experiment has been covered in detail in the literature~\cite{M.Dueren1995}--\cite{CollaborationNucl.Instrum.Meth.A540:68-1012005} and no such detailed description will be repeated here. This work will refer solely to BH/DVCS detected by scattering a positron or electron off a proton target, with the produced photons being detected in the electromagnetic calorimeter~\cite{Avakian1998}. Bethe Heitler/Deeply Virtual Compton Scattering events are selected in the present H{\sc ermes} analysis by use of a missing mass technique --- the recoiling proton is scattered outside of the H{\sc ermes} acceptance, so the final data sample comprises the desired BH/DVCS candidate events and some background processes~\cite{HERMESCollaborationPhys.Rev.Lett.87:1820012001}--\cite{HERMEScollaboration2010}.
 The most egregious of these background processes is BH/DVCS involving a proton resonance which typically makes up approx. 10\% of the data sample. Although the H{\sc ermes} spectrometer was upgraded in 2006 with a recoil proton detector in the target region~\cite{KaiserR.etal2001} in order to eliminate this background process, the current technique does not make use of it and it will not be discussed here further.

\section{Deeply Virtual Compton Scattering}

H{\sc ermes} was designed to measure asymmetries. Asymmetries in the azimuthal distribution of the produced real photon in DVCS are measured to provide information that can be used to constrain Generalised Parton Distributions~\cite{Belitsky2002}. As a consequence of the unique H{\sc ermes} setup, there are several different combinations of beam and target states that can supply useful information for DVCS:
\begin{eqnarray}
\mathcal{A}_{\rm{C}} = \frac{\rm{d}\sigma^+ - \rm{d}\sigma^-} {\rm{d}\sigma^+ + \rm{d}\sigma^-} \label{e:bca}\\
\mathcal{A}_{\rm{LU}} = \frac{\rm{d}\sigma^\rightarrow - \rm{d}\sigma^\leftarrow}{\rm{d}\sigma^\rightarrow + \rm{d}\sigma^\leftarrow}\label{e:bsa}\\
\mathcal{A}_{\rm{UL}} = \frac{\rm{d}\sigma^\Rightarrow - \rm{d}\sigma^\Leftarrow}{\rm{d}\sigma^\Rightarrow + \rm{d}\sigma^\Leftarrow}\label{e:aul}\\
\mathcal{A}_{\rm{LL}} = \frac{\rm{d}\sigma^{\stackrel{\rightarrow}{\Rightarrow}} - \rm{d}\sigma^{\stackrel{\leftarrow}{\Leftarrow}}}{\rm{d}\sigma^{\stackrel{\rightarrow}{\Rightarrow}} + \rm{d}\sigma^{\stackrel{\leftarrow}{\Leftarrow}}}\label{e:all}
\end{eqnarray}
Here $\rightarrow$ ($\leftarrow$) refers to beam helicity states and $\Rightarrow$ ($\Leftarrow$) refers to target helicity states. The beam charge asymmetry in Eq.~\ref{e:bca} is expected to be sensitive mostly to the real part of CFF $\mathcal{H}$, the beam helicity asymmetry in Eq.~\ref{e:bsa} is expected to be sensitive mostly to the imaginary part of CFF $\mathcal{H}$, the longitudinal target spin asymmetry in Eq.~\ref{e:aul} provides access to the imaginary part of $\mathcal{\widetilde{H}}$ and the double spin asymmetry in Eq.~\ref{e:all} provides access to the real part of $\mathcal{\widetilde{H}}$. Since H{\sc era} supplies both beam charges, the interference and DVCS contributions to Eq.~\ref{e:bsa} can be extracted separately.

There are currently two approaches for obtaining GPD information from experimental data. One approach is to fit the CFFs from experimental measurements of asymmetries simultaneously, thus revealing information on the CFFs but not the underlying GPDs~\cite{H.Moutarde2009}. This approach has the advantage that it is fast and easily understandable, requiring no detailed theoretical work on the underlying GPD, but provides detail only on the CFF which cannot be used to extract underlying physical quantities. The second more thorough approach is to postulate GPDs from first principles and work through the detailed calculations to produce predictions for asymmetries. The underlying calculations can then be revised to provide predictions that are more consistent with the observed data~\cite{Kumericki}~\cite{Kumericki2008}.

\section{DVCS Measurements @ H{\sc ermes}}

\begin{figure}
\includegraphics[width=0.8\textwidth]{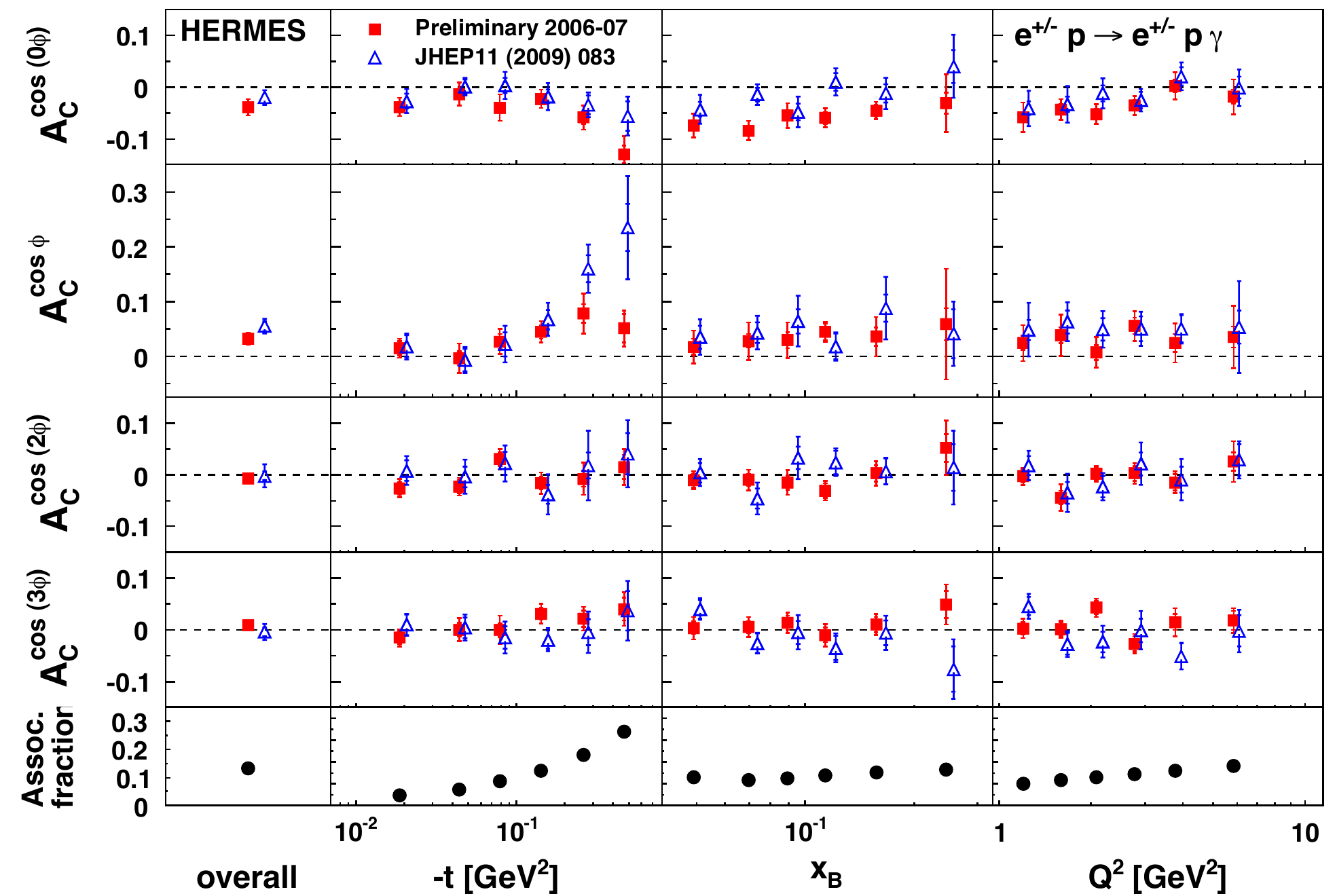}
\caption{The Beam Charge Asymmetry measured at H{\sc ermes}. See text for details.}
\label{f:bca}
\end{figure}

\begin{figure}
\includegraphics[width=0.8\textwidth]{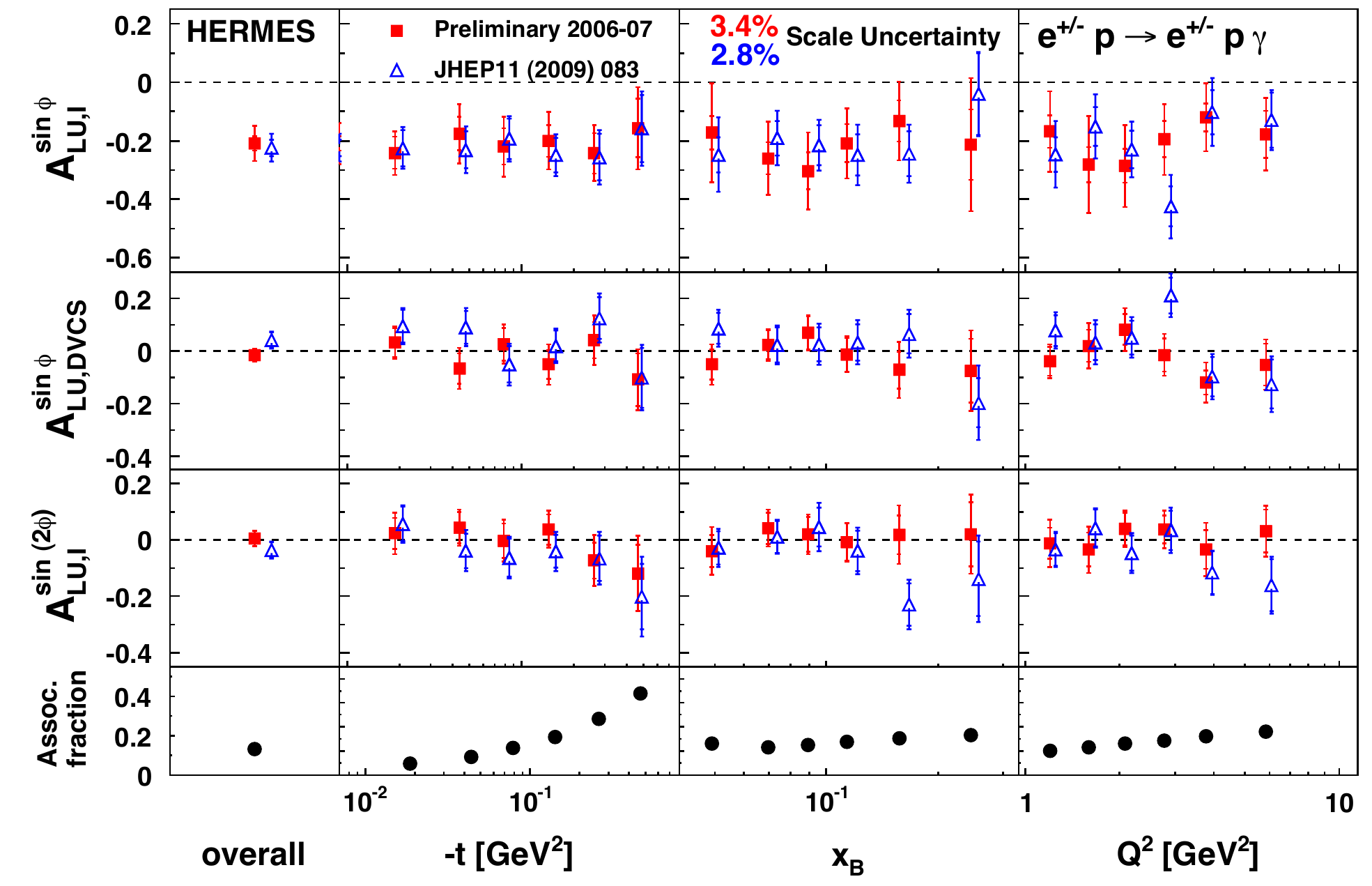}
\caption{The Beam Helicity Asymmetry measured at H{\sc ermes}. See text for details.}
\label{f:bsa}
\end{figure}

A selection of the DVCS measurements made at H{\sc ermes} are presented in Figs.~\ref{f:bca} and~\ref{f:bsa}, corresponding to the asymmetries in Eqs.~\ref{e:bca} and~\ref{e:bsa}~\cite{2009a}~\cite{Burnsth2010}. The lower panel in each of the figures shows the expected contamination in the data sample from resonance events which have an unknown effect on the extracted asymmetries. Asymmetries with a leading twist contribution from GPD $H$ are given by the $\cos\phi$ harmonic in Fig.~\ref{f:bca} and the $\sin\phi$ harmonic in Fig.~\ref{f:bsa}. The other harmonics in the figures show sub-leading twist contributions or harmonics that are expected to be significantly suppressed at H{\sc ermes} kinematics.

The two sets of data in Figs.~\ref{f:bca} and~\ref{f:bsa} represent independent agglomerations of the H{\sc ermes} data productions. The square points represent data from 2006-07 extracted using the traditional H{\sc ermes} analysis technique. The triangles points are data extracted from H{\sc ermes} data taken between 1996 and 2005. The two data sets broadly agree with each other across the entire range of harmonics, showing that the leading-twist contributions give, as expected, the strongest signal in the data. Extractions of the asymmetry amplitudes of Eqs.~\ref{e:aul} and~\ref{e:all} have also been published by H{\sc ermes}~\cite{HERMEScollaboration2010}~\cite{Mahon2010}.
\clearpage

\bibliographystyle{aipproc}
\bibliography{jburnsref}

\begin{thebibliography}{27}
\expandafter\ifx\csname natexlab\endcsname\relax\def\natexlab#1{#1}\fi
\providecommand{\enquote}[1]{``#1''}
\expandafter\ifx\csname url\endcsname\relax
  \def\url#1{\texttt{#1}}\fi
\expandafter\ifx\csname urlprefix\endcsname\relax\def\urlprefix{URL }\fi
\providecommand{\eprint}[2][]{\url{#2}}

\bibitem[Ji(1997)]{Ji1997}
X.~Ji, \emph{Phys. Rev. D} \textbf{55}, 7114--7125 (1997).

\bibitem[Burkardt(2002)]{BurkardtPhys.Rev.D62:0715032000;Erratum-ibid.D66:1199%
032002}
M.~Burkardt, \emph{Phys. Rev. D} \textbf{66}, 114005 (2002).

\bibitem[Balitsky and Radyushkin(1997)]{Balitsky1997}
I.~I. Balitsky, and A.~V. Radyushkin, \emph{Phys. Lett. B} \textbf{413},
  114--121 (1997).

\bibitem[Dueren(1995)]{M.Dueren1995}
M.~Dueren, \emph{Universit\"at Erlangen-Nuernberg, DESY Grey Report:
  DESY-HERMES 95-02}  (1995).

\bibitem[{Ackerstaff et al.}(1998)]{Ackerstaff1998}
K.~{Ackerstaff et al.}, \emph{Nucl. Instrum. Meth. A} \textbf{417}, 230--265
  (1998).

\bibitem[{Rykbosch et~al.}(1999)]{Rykbashetal.1999}
D.~{Rykbosch et~al.}, \emph{Nucl. Instrum Meth. A} \textbf{433}, 98 (1999).

\bibitem[{Benisch et al.}(2001)]{2001}
T.~{Benisch et al.}, \emph{Nucl. Instrum. Meth. A} \textbf{471}, 314--324
  (2001).

\bibitem[{Beckmann et al.}(2002)]{Beckmann2002}
M.~{Beckmann et al.}, \emph{Nucl. Instrum. Meth. A} \textbf{479}, 334--348
  (2002).

\bibitem[A.{Nass et~al.}(2003)]{2003}
A.{Nass et~al.}, \emph{Nucl. Instrum. Meth. A} \textbf{505}, 633--644 (2003).

\bibitem[{Baumgarten et al.}(2003{\natexlab{a}})]{2003a}
C.~{Baumgarten et al.}, \emph{Nucl. Instrum. Meth. A} \textbf{508}, 268--275
  (2003{\natexlab{a}}).

\bibitem[{Baumgarten et al.}(2003{\natexlab{b}})]{2003b}
C.~{Baumgarten et al.}, \emph{Nucl. Instrum. Meth. A} \textbf{496}, 277--285
  (2003{\natexlab{b}}).

\bibitem[{The HERMES
  Collaboration}(2005)]{CollaborationNucl.Instrum.Meth.A540:68-1012005}
{The HERMES Collaboration}, \emph{Nucl. Instrum. Meth. A} \textbf{540}, 68--101
  (2005).

\bibitem[{Avakian et al.}(1998)]{Avakian1998}
H.~{Avakian et al.}, \emph{Nucl. Instrum. Meth. A} \textbf{417}, 69--78 (1998).

\bibitem[{The HERMES
  Collaboration.}(2001)]{HERMESCollaborationPhys.Rev.Lett.87:1820012001}
{The HERMES Collaboration.}, \emph{Phys. Rev. Lett.} \textbf{87}, 182001
  (2001).

\bibitem[{The HERMES
  Collaboration}(2007)]{CollaborationPhys.Rev.D75:0111032007}
{The HERMES Collaboration}, \emph{Phys. Rev. D} \textbf{75}, 011103 (2007).

\bibitem[{The HERMES Collaboration}(2008)]{HERMESCollaboration2008}
{The HERMES Collaboration}, \emph{JHEP} \textbf{06}, 066 (2008).

\bibitem[{The HERMES Collaboration}(2009{\natexlab{a}})]{2009a}
{The HERMES Collaboration}, \emph{JHEP} \textbf{0911}, 083
  (2009{\natexlab{a}}).

\bibitem[{The HERMES Collaboration}(2010{\natexlab{a}})]{nuclmass2009}
{The HERMES Collaboration}, \emph{Phys. Rev. C} \textbf{81}, 035202
  (2010{\natexlab{a}}).

\bibitem[{The HERMES Collaboration}(2009{\natexlab{b}})]{2010}
{The HERMES Collaboration}, \emph{Nucl. Phys. B} \textbf{829}, 1--27
  (2009{\natexlab{b}}).

\bibitem[{The HERMES
  Collaboration}(2010{\natexlab{b}})]{HERMEScollaboration2010}
{The HERMES Collaboration}, \emph{JHEP} \textbf{06} (2010{\natexlab{b}}).

\bibitem[{Kaiser et al.}(2001)]{KaiserR.etal2001}
R.~{Kaiser et al.}, \emph{DESY-PRC} \textbf{97-06-ADD}, 36 (2001).

\bibitem[Belitsky et~al.(2002)]{Belitsky2002}
A.~V. Belitsky, D.~Mueller, and A.~Kirchner, \emph{Nucl. Phys. B} \textbf{629},
  323--392 (2002).

\bibitem[H.Moutarde(2009)]{H.Moutarde2009}
H.Moutarde, \emph{Phys. Rev. D} \textbf{79}, 094021 (2009).

\bibitem[Kumericki and D.Mueller(2010)]{Kumericki}
K.~Kumericki, and D.Mueller, \emph{Nucl. Phys. B} \textbf{841}, 1--58 (2010).

\bibitem[Kumericki et~al.(2008)]{Kumericki2008}
K.~Kumericki, D.~Mueller, and K.~Passek-Kumericki, \emph{Proceedings of DIS
  2008}  (2008).

\bibitem[Burns(2010)]{Burnsth2010}
J.~R. Burns, \emph{Deeply {V}irtual {C}ompton {S}cattering off an {U}npolarised
  {H}ydrogen Target at {HERMES}}, Ph.D. thesis, University of Glasgow (2010).

\bibitem[Mahon(2010)]{Mahon2010}
D.~F. Mahon, \emph{Deeply {V}irtual {C}ompton {S}cattering off {L}ongitudinally
  {P}olarised {P}rotons at {HERMES}}, Ph.D. thesis, University of Glasgow
  (2010).

\end{thebibliography}
\end{document}